Accurate Target Localization by using Artificial Pinnae of brown long-eared bat

**Authors**

Sen Zhang[1], Xin Ma[1,2]*, Hongwang Lu[3], Weikai He[4], Weidong Zhou[5]

**Affiliations**

1 School of Information Science and Engineering, Shandong University, Qingdao 266237, China
2 Shenzhen Research Institute, Shandong University, Shenzhen 518000, China
3 School of Physics, Shandong University, Jinan 250022, China
4 School of Physics, University of Jinan, Jinan 250000, Shandong, China
5 School of Microelectronics, Shandong University, Jinan 250022, China

Abstract

Echolocating bats locate the targets by echolocation. Many theoretical frameworks have been suggested the abilities of bats are related to the shapes of bats ears, but few artificial bat-like ears have been made to mimic the abilities, the difficulty of which lies in the determination of the elevation angle of the target. In this study, we present a device with artificial bat pinnae modeling by the ears of brown long-eared bat (Plecotus auritus) which can accurately estimate the elevation angle of the aerial target by virtue of active sonar. An artificial neural-network with the labeled data obtained from echoes as the trained and tested data is used and optimized by a tenfold cross-validation technique. A decision method we named sliding window averaging algorithm is designed for getting the estimation results of elevation. At last, a right-angle pinnae construction is designed for determining direction of the target. The results show a higher accuracy for the direction determination of the single target. The results also demonstrate that for the Plecotus auritus bat, not only the binaural shapes, but the binaural relative orientations also play important roles in the target localization.

**MAIN TEXT**

**Introduction**

The rapid development of mobile robots and flying robots emphasizes the need for new sensory approaches for target search and obstacle avoidance. Although the machine vision can meet the basic requirements in most cases, there are always some situations, such as darkness, smoke, which are not suitable for the application of vision. Sonar sensing is an effective complement to the vision sensing in robot applications. Most bats can use sonar to navigate and forage or hunt for insects[1-3], so they are also often called echolocating bats [4-8].The sonar used in echolocating bats enables them to move flexibly in dark and complex environments , determines the existing of targets in space [9] and obtain location information [10].Some of the echolocating bats can discriminate the target's angular resolution up to 1.5° [11].These performances of the echolocating bats have drawn great attention in the research field.

Many researchers have focused on the study of the bats sonar and great progress has been made in the mechanism of bat's echolocation [12,13]. Extensive studies have shown the spatial information of targets can be obtained by analyzing the received echoes [14,15].Several studies have display the detail application of the bat-like echoes in navigation or target localization. For example, Yamada Y et al. designed a vehicle that could automatically avoid obstacles mimicking bat sonar behavior [16]. Dieter Vanderelst et al. conducted the scene recognition experiments by using ensonification data [17]. Itamar Eliakim presented an autonomous terrestrial robot that can map a novel environment relying on echolocation data classification[18]. But we can note these bat inspired sonar designs are quite different from real artificial bat sonar, because they lack the pinna construction of the bat ears which can play useful roles in bat's target localization[19]. Also, most of the above designs are 2D sonars realized mainly by microphone arrays time delay estimation, while the true artificial echolocating bat sonar should be able to localize realistic targets in 3D space with a binaural bio-inspired sonar design. There are also a few research results that are closer to the echolocating bats. Müller et al.[20] theoretically gives the possibility of accurately locating by artificial bats in 3D. Schillebeeckx et al. [21] have designed a binaural artificial bat-like head for localizing realistic targets and the result shows the feasibility of targets location in 3D space.

The substance of the most of reports in bat inspired target localization designs is obtaining the azimuth angle or the distance determination based on interaural time difference (ITD) cues [22,23] ,while in the 3D space localization, the elevation estimation of the target is important whether for navigation or target localization. There has not been much research on elevation features of bat ears structures, nor has complex information contained in echoes received by the bat ears been fully utilized. It is best if we can find a configuration which has good directional sensitivity in the elevation while insensitive to the azimuth. Many reports have suggested that some physiological structures have evolved which are related to the bat's echolocation and space directivity [24][25][26]. Among these cases，the pinnae structure of brown long-eared bat (Plecotus auritus) pinnae has its special characteristics [27，28]: the side lobe of the acoustic beam formed by the pinna sweeps a spatial region with the change of frequency of the sonar, and our previous study[29] show the peak of the spatial frequency response shifts linearly with the elevation angle. Inspired by these characteristics, we have mimicked this approach in the sonar device: we employ artificial 3D printed brown long-eared bat's pinnae with one microphone in the root of each pinna to mimick bat's ears, and use an ultrasonic speaker mimicking the bat's mouth which produced frequency modulated (FM) chirps pulse train as a typical bat, reproduce the discrimination in elevation direction. Before the work of artificial ears realizing the echolocation, first we conducted a more detailed simulation experiment than the description in our previous work by using the finite element method (FEM) [30] and Kirchhoff integral [31] which supply us more enough material for our artificial bat like work, then the bat-like artificial ears echolocation experiments were conducted and the results demonstrate under the condition that the two pinnae point to the same direction, active head-related spectral features of the echoes have good cluster effects with the elevation, while poor with the azimuth. Based on the correlation between the spectral features of the echoes and the target elevation, we succeed in estimating a target angle (azimuth $\theta$ and elevation $\varphi$) by making the two pinnae perpendicular to each other.

**Results**

Our goal is to realize the accurate echolocation for single target by using bat-like artificial device, The results show when the directions of the two pinnae are to parallel, our approach achieved high estimation accuracy in elevation direction; and when the two pinnae in the device are perpendicular to each other, it can be used for estimating a target angle (azimuth $\theta$ and elevation $\varphi$).

**1 Spacial Frequency characteristics of the digital Brown big-eared bat's pinna**

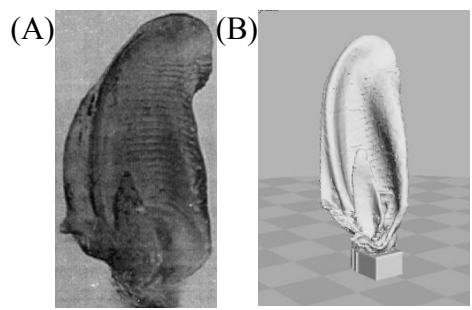

**Fig. 1. Photos and 3D model of *Plecotus auritus* ear.**

**(A)** 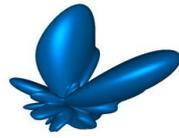
33.5kHz

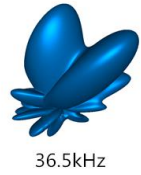
36.5kHz

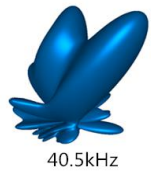
40.5kHz

**(B)** 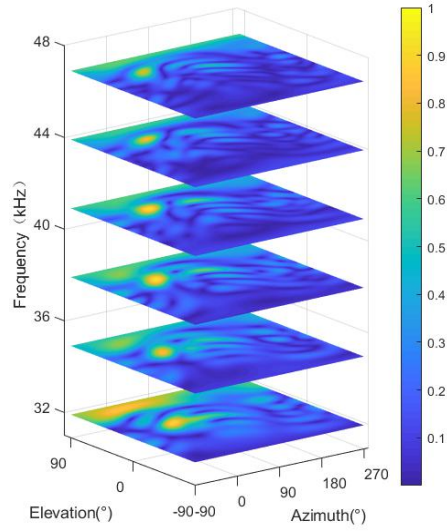

（C） 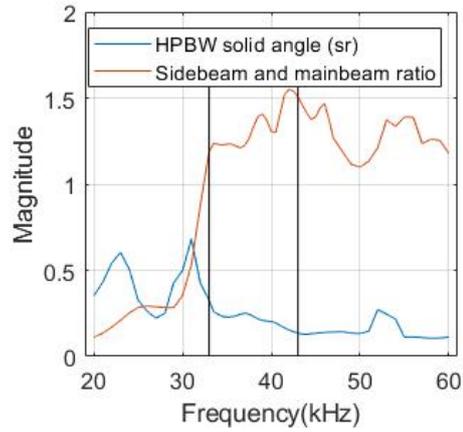

（D） 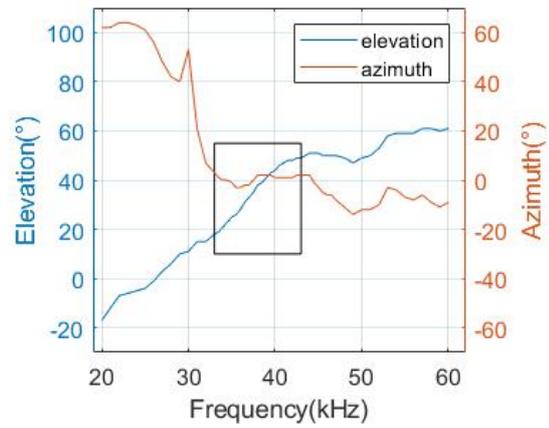

(E) 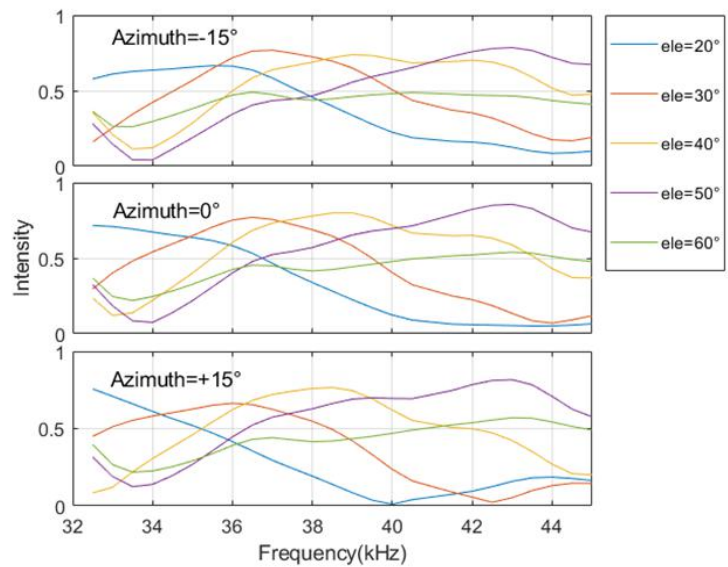

**Fig. 2. Bat ear model far-field pattern. (A)** Distance field lobe pattern at different frequencies (33.5 kHz, 36.5 kHz and 40.5 kHz) from up to down. **(B)** Raw data of acoustic far field obtained using Kirchhoff integral. With change in frequency, the size and directivity of the side lobe changed in elevation. **(C)** Half-power beam width and energy ratio of side lobe and main lobe. **(D)** Side lobe elevation and azimuth. **(E)** Frequency response curve of fixed azimuth with elevation range from 20° to 60° (every 10°). The frequency response curve at different elevation presents different peaks because of frequency scanning characteristic. When frequency changes, the lobe under this frequency points to different directions.

Frequencies spanning the entire frequency range (22 to 56 kHz in 1kHz steps) known to be covered by the biosonar pulses' first harmonic [32] were analyzed by using numeral method (see Methods). The first side lobe in the beam pattern (Fig. 2A,B) performs a frequency-driven scanning characteristic and a relatively strong power when the frequencies exceed 32.5 kHz. The half-power beam width (HPBW) curve is relatively large while the power of side beam is low when the frequencies are less than 30kHz (Fig. 2C). When the frequencies are less than 30kHz, the directivity of the lobe under these frequencies is not concentrated which will cause low resolution. The beam direction of the first side lobe shifts along the elevation almost linearly with the change of frequency in the band 30KHz to 60 KHz while azimuth of the side lobes almost maintain stable (Fig. 2D), which suggests in this frequency band only the elevation information have strong correlation.

The corresponding location of the peak in the frequency response shifts with the different elevations because of frequency scanning characteristics of the first side lobe (Fig. 2E). This suggests that the elevations information of a target located can be obtained by measuring the frequency response of the sound emitted or reflected by the target. According to this feature, bats' received echoes carry different features that can be used to distinguish elevation of the target, which is relevant for this work. We conducted an actual positioning and extracted features of different frequencies from the resulting echo to locate elevation.

**2 Target echolocation by artificial bat-like sonar device**

Although the special information generated by virtue of scanning sidelobes have been demonstrated in the work of simulations, the physical verification need to be done for further ascertainment of the function, and then we have designed a bat-like device to determine spatial location for the point-like target (Fig3.A) by taking advantage of the function like the *Plecotus auritus* ears.

## 2.1 The target angle estimation in the case of parallel pinnae

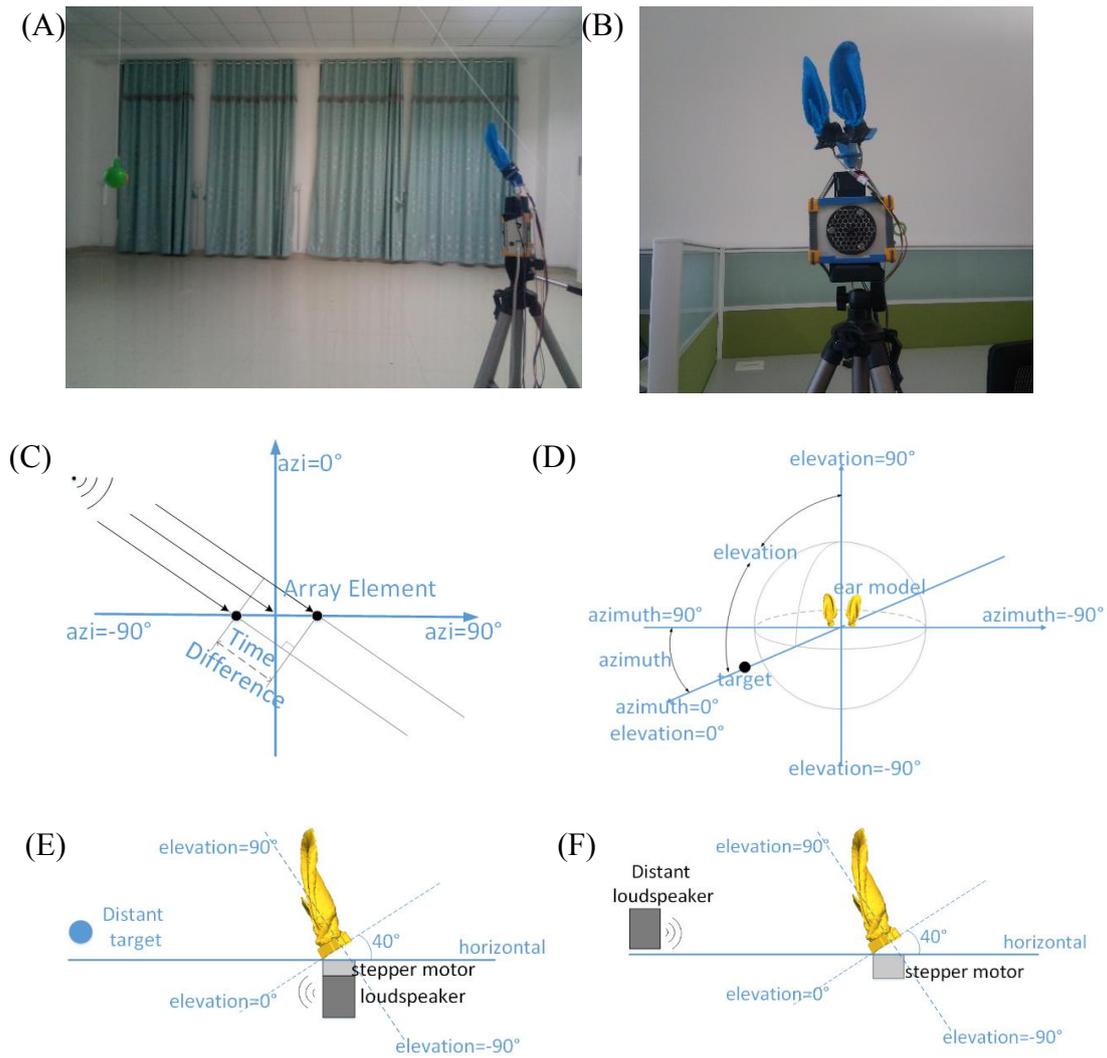

**Fig. 3. Bats sounder acquisition geometry scheme. The whole collection device is respectively bat ear model, ultrasonic microphone, stepping motor, ultrasonic loudspeaker and support frame from top to bottom. (A, B) Actual environment (C) Top view of platform. (D) Spherical coordinate. (E, F) Side view of the artificial pinna.**

In this section, first, we developed the biomimetric sonar device in which the of the two artificial pinnae point to the same orientations. Then we used the artificial bat-like sonar device to emit and receive signals and collect the data according to table 1. Inspired by special location remembering abilities of the bats  [33], machine learning and statistical methods were adopted. In the large of learning methods, BP neural network is comparatively simple and have good effects in a wide scope of applications. Based on this, in this paper, a BP neural network was trained for the task of target angle identification.

**Table 1. Experimental parameters.**

| ID | Name | Comment |
|---|---|---|
| 1 | Sampling rate | 100 kHz |
| 2 | Model size of bat ear | 3 times the size of the original size |
| 3 | Microphone distance | 5 cm |
| 4 | relative position between target and device | 8 different locations in section 2.1 and training data acquition of section 2.2.(see Fig.10). Random selected position 1.0-2.0 meters from target in test data acquition of section 2.2. |
| 5 | Transmission signal | Linear frequency modulation signal |
| 6 | Frequency range | 5–20 kHz |
| 7 | Passive signal duration | 2 s |
| 8 | Active signal duration | 5 ms |
| 9 | Azimuth rotation angle | -90°–90°(step length: 7°) in section 2.1, -28°– 28°(step length: 7°) in section 2.2. |
| 10 | Elevation rotation angle | 20°–55° (step length: 5°) in section 2.1, 12–68°(step length: 7°)in section 2.2. |

2.1.1 Single pulse target elevation estimation

A set of time-frequencies representations were extracted from the echoes and used as input for the network (Methods, Fig. 4A). For testing the estimation accuracy of elevation under different scope of azimuth, in this section, the single pulse and different sets of data categories in every statistics (Methods, and Fig. 4C) are adopted, the training data and the testing data are different but come from the same dataset depicted in Method. Ten-fold cross validation were conducted for getting more credible results, the means of the results of ten-fold cross validation are shown in Fig. 4B.

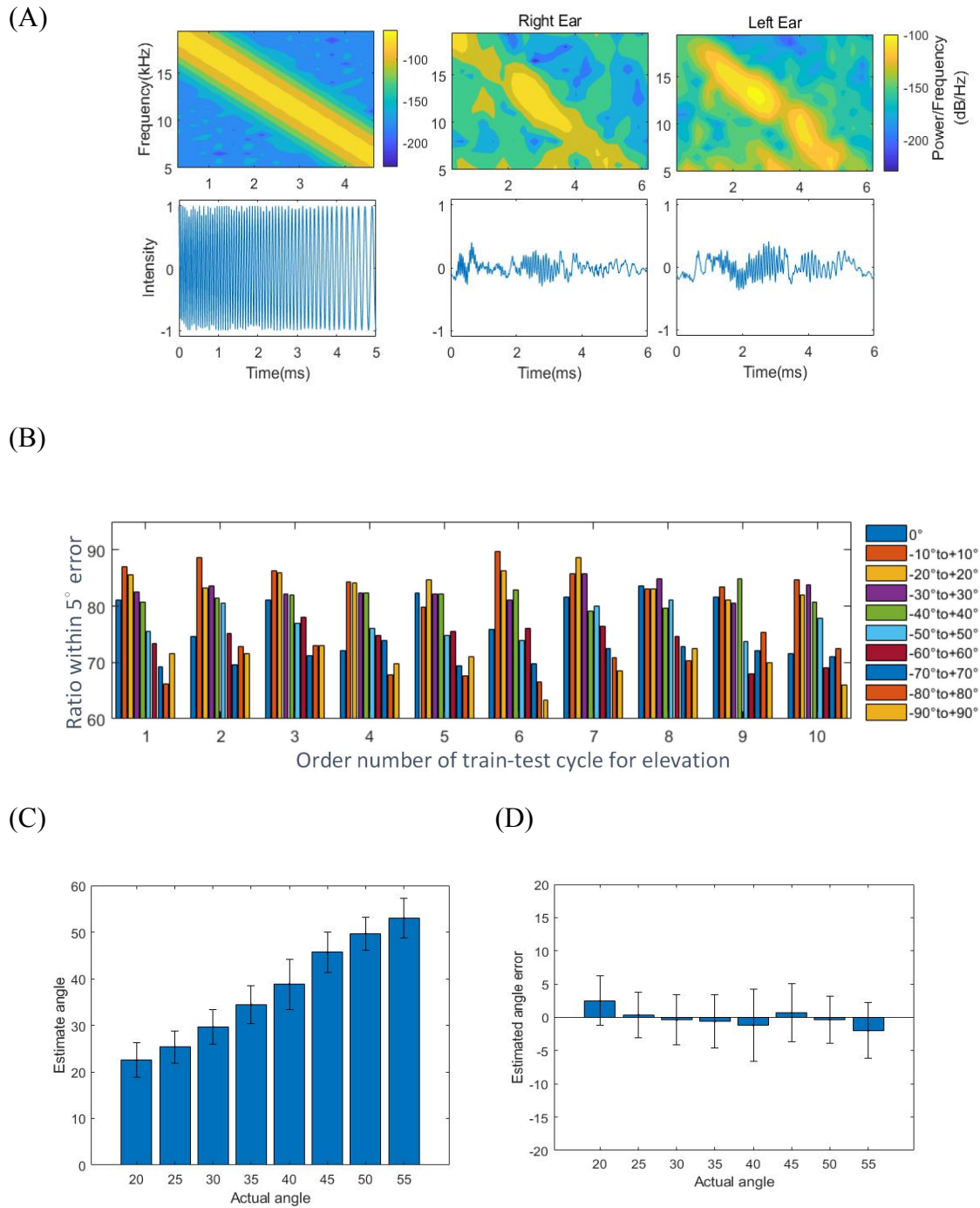

**Fig. 4. Chirp pulse, echos and the results of neural network recognition. (A)** An example of an emitting signal by the loudspeaker and echoes received by the two ears. **(B)** Ratio under ±5° error in 10 cross-validation cycles. 10 histograms in each cycle represent the limit scope of azimuth of training set from ±0° to ±90°respectively. The vertical axis represents the ratio of error within 5° between estimated angle and actual angle of elevation. **(C)** the difference between true elevation and estimated elevation. The top level of blue histogram stands for the mean of the estimation values, and black bar in the blue histogram stands for the error distribution, The neural network predicts give the best result when the

**limiting angle of azimuth is concentrated close to 0°. When the scope of azimuth increased, and estimation accuracy will declined. (D) Correspondence between true elevation and elevation error. The closer the average angle error was to zero and the smaller the standard deviation of error bar, the better the training result of neural network.**

The limit angle of azimuth was in the range of -$N°$ to +$N°$ which stands for the left and the right (see Fig.3). The best elevation estimation emerged within the range ±30° of azimuth, with the ratio of estimation value with less than 5° error exceeding 80%.

This result is largely consistent with our simulation results. The far-field direction diagram of the bionic bat ear beam pattern (Fig.2) indicates that the half-power wave lobe width of the main lobe is oriented towards a zero-azimuth angle with a change of approximately 20°. This result is also consistent with echolocation function of actual bats. When bat ears are facing the target, the function of positioning is most powerful. When not facing the target, bat can ensure the correct orientation angle by rotating its head and ears to face the target.

With the increase of the azimuth range of participants in training, the estimation accuracy of the elevation will decline but still remains above 60% (fig 4 B)，this mainly because the orientation of the transmitter we used remains in the 0° of the aspects which caused the signal-to-noise ratio (SNR) is much lower in the high azimuth than in the low azimuth. But the standard deviation of the elevation results within 20-55° almost remains stable, this indicate although the SNR becomes poor in high elevation like in the azimuth, the correlations between sweeping frequency and the elevation are strongly maintained as long as the range of the elevation is not too large.

2.1.2 Improving the echolocation accuracy by using Pulse train

Up to now we have obtained the estimation value of elevation from single pulse. For imitating the pulse train used by the bats, pulse train estimation method was used and moving windows accumulation method (see Methods) was designed for getting a more accurate result.

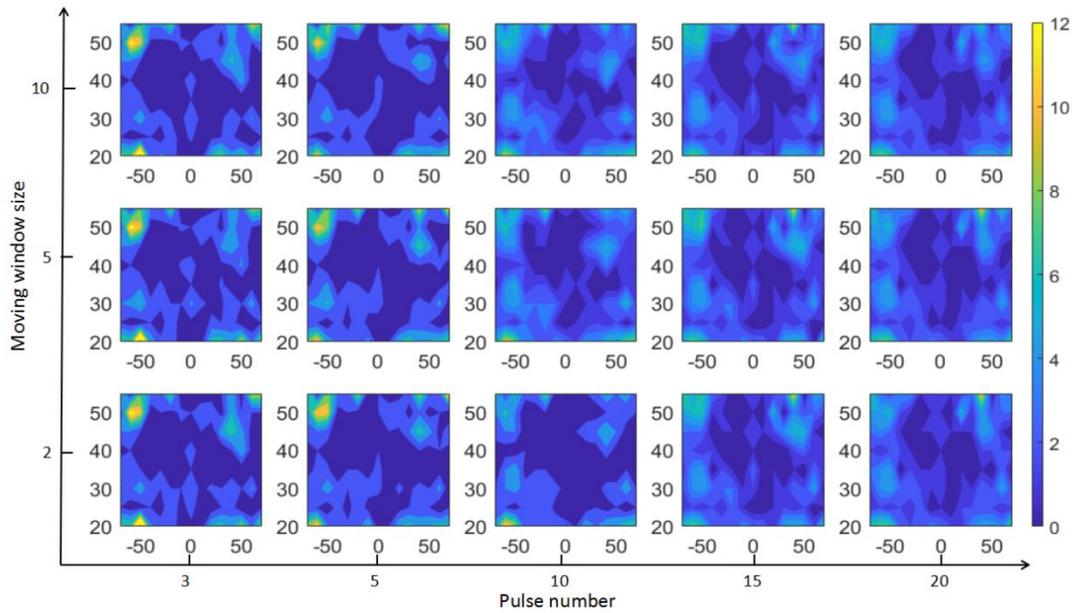

Fig. 5. The error between the estimated elevation and true elevation by pulse chain estimation, under the condition of different size of pulse train (from left to right the number of pulses in a single pulse train is 3 to 20) and different level of moving windows(from top to the bottom the size of moving window size is 10, 5, 2). The horizontal and vertical axis in each small diagram represent limited azimuth used in training set(in this figure,the range is ±70° only) and the estimated elevation (the range is 20° to 55°)respectively.

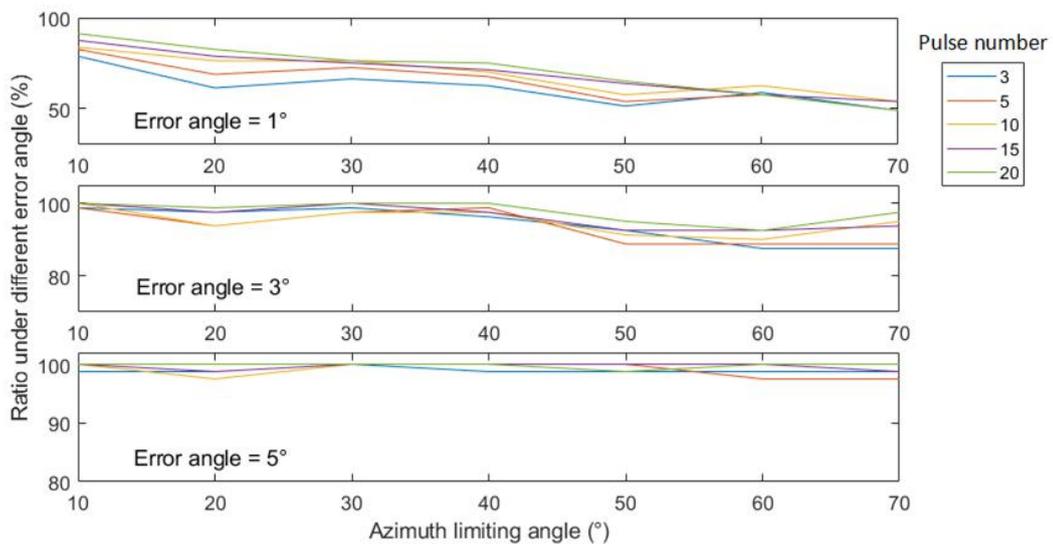

Fig. 6. Pulse train based accuracy of the elevation estimation with different error range (From top to bottom: ±1°, ±3°, and ±5° ) under different limit range of azimuth. The different colors stand for different number of pulses in one pulse train used for estimation.

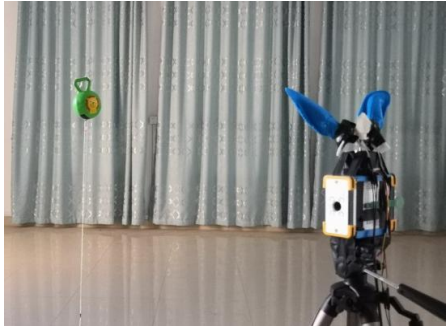
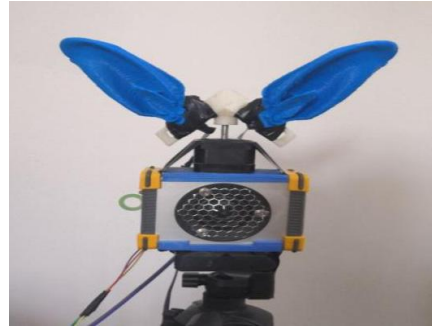

(a)

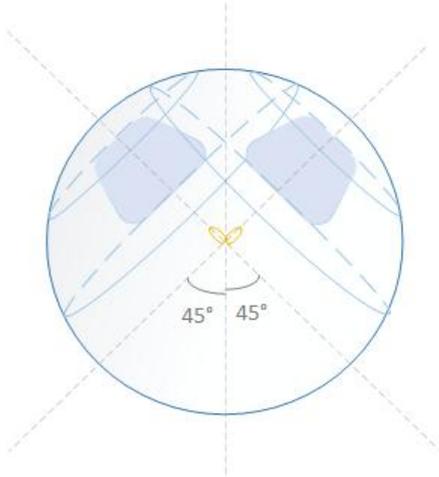
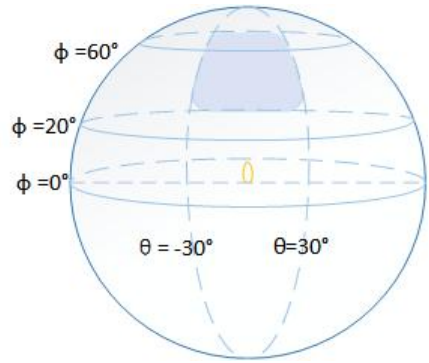

(b)                                      (c)

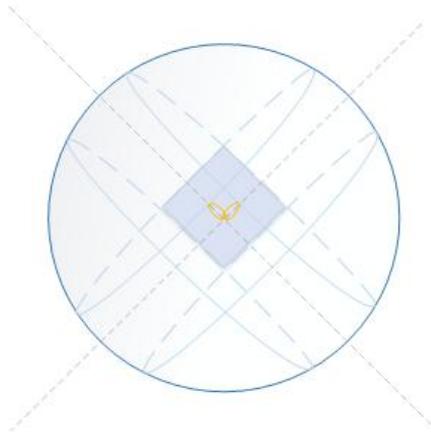
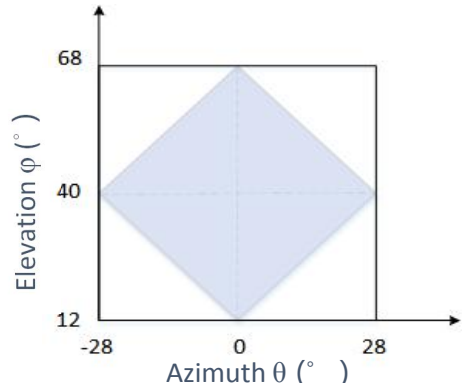

(d)                                      (e)

Fig 7. **The orthogonal pinnae active sonar device. Every pinna tilts forward 40 degrees, so as to form the overlap region(shadow in fig11) of the two pinna effective ranges for their respective elevation estimation. The shadow region represent overlap of the two pinna effective ranges for their respective elevation estimation. The actual statistical scope is -28° to 28° in azimuth and 12~68° in elevation as fig 7 (e).**

The increase of numbers of pulses in pulse train effectively compensates for the loss of elevation estimation caused by the larger range of azimuth (middle and the bottom in Fig.6).when the number of pulses in one pulse train attained to 20,the elevation estimation accuracy can achieve an accuracy of more than 95% with an error range of ±3° error (the middle in Fig.6).

## 2.2 The target angle estimation in the case of orthogonal pinnae

In the process of estimation of elevation described in above sections, the two pinnae of the big brown bat are parallel to each other. We have observed that the two pinnae of the big brown bats often stretch to a certain angle when foraging. Obviously if the angle is 90, the orthogonality of the two pinnae can be utilized to obtain the aspect angles in the two orthogonal directions.

The artificial big brown bat pinnae used in target localization are shown in Fig.7. Apart from the angle of the two pinnae, the other measurement conditions are same with the elevation estimation depicted in section 2.1.

The training samples were obtained as illustrated in fig 10.(see Method).The total number of ehoes pulses is 65600 (In each of the eight sites, the orthogonal pinnae active sonar device collects 200 echo pulses from each of the 41 aspect direction ). The test data are different from the test data used in above tests. For observing the robustness of the system used for estimation, Robustness tests methods are designed to verify their generalization abilities. Unlike the data set used in cross validation depicted in section 2.2 and the Method, the robustness testing samples are collected in the same experimental chamber but under the condition that the acquisition device (include the right-angle pinnae) were placed almost in the random directions relative to the target (small ball), and the distance between the pinnae and the ball is almost random selected within the range of 1.0m to 2m.the whole process is performed like section2.2, the azimuth and the elevation angle are independently clustered by BP neural network with 9 neurons in the hidden layer, and the moving windows accumulation method (see Method) is also used for getting the final estimation values.

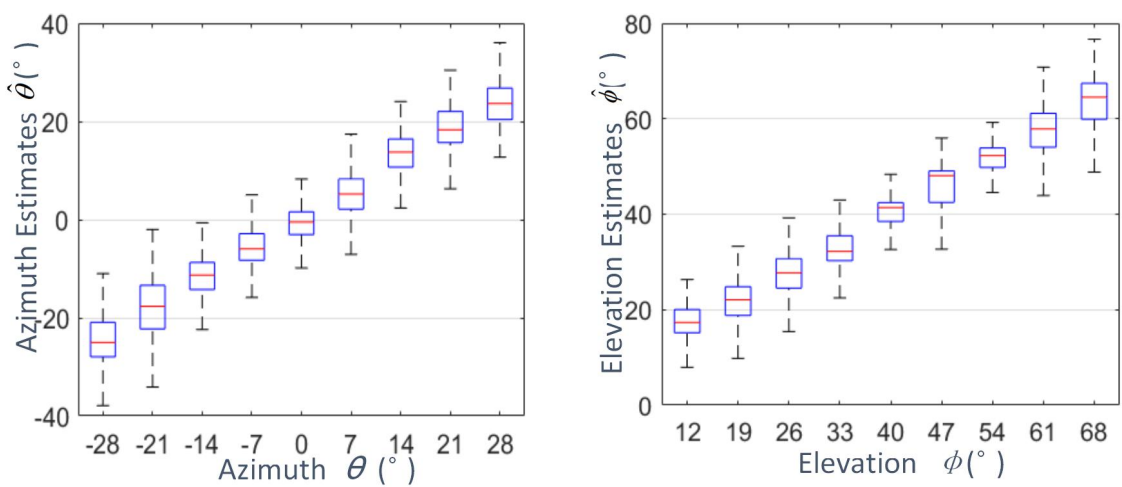

**Fig.8.The statistics of the azimuth estimation (a) and the elevation estimation**

**(b). in the single pulse test.**

The results shown in Fig.8 demonstrated that not only for the elevation estimation but for the azimuth estimation can be successfully fulfilled. For the azimuth estimation, the estimation value is close to the mean in the adjacency of zero-degree angle. The farther the deviation from 0 degree in the negative direction is, the greater the negative deviation of the estimated value is, and the opposite is true in the positive direction. For the elevation estimation, the estimation value is close to the mean in the adjacency of 33 degree angle, the farther the deviation from the 33 degree in the negative direction is, the greater the negative deviation of the estimated value is, and the opposite is true in the positive direction.

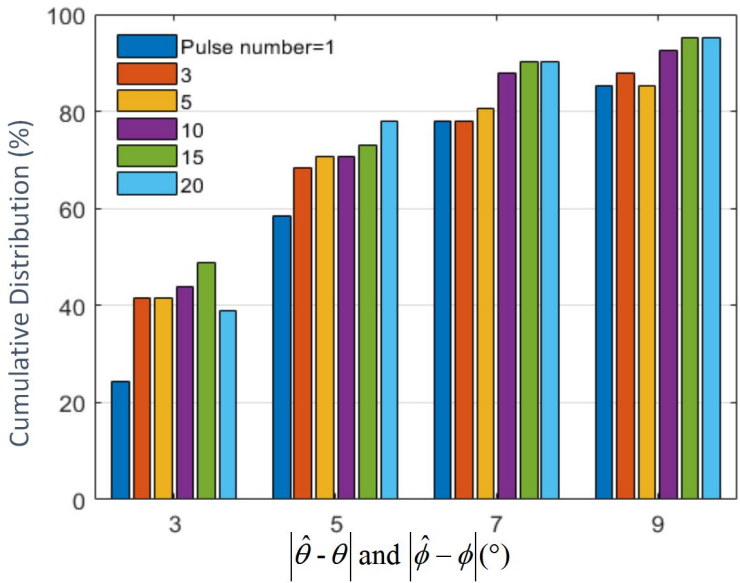

**Fig.9. Cumulative distributions for the corresponding azimuth and elevation errors. The abscissa axis represents the errors that both the azimuth and elevation need meet. In our experiments.**

Figure 9 shows the localization results under different pulse train. For the all pulse train, under 50% of all target angles are estimated, with an error $|\hat{\theta} - \theta| \leq 3°$ and $|\hat{\phi} - \phi| \leq 3°$, and under 91% with an error $|\hat{\theta} - \theta| \leq 6°$ and $|\hat{\phi} - \phi| \leq 6°$. These percentages increase as the number of pulse in these pulse trains rise. Only less than 5% error rate emerge when the demand is $|\hat{\theta} - \theta| \leq 9°$ and $|\hat{\phi} - \phi| \leq 9°$ under the condition that number of pulses in each pulse train equals or exceeds 15.

Obviously, using pulse train can improve the estimation results. Nonetheless, the fig.9 shows not that the more the number of pulses is included in each pulse train, the better the estimation effect is. The optimal effect appears when the number of pulses in each pulse train is 10~15.

Compared with the results in section2.1, the results look not so good. At a rough glance, the results don't seem to be so good Compared with the results in section2.1.but in fact, they have different concepts which were shown mainly in two aspect: One is the result here is the location estimation results, while the results in section 2.1 is the results of elevation estimation. Another is the difference in the test condition. In section 2.1, the bat-like device was located in the fixed eight locations surrounding the target, with a fixed distance of 1.5m to the target, but the bat-like device here was located in random locations with a distance of 1-2m to the target. So the test results here own more strong generalization ability, and as the diameter of the target (small ball) is 11 cm, of which the relatively excessive size may cause the lower angular resolution, these results indicate an acceptable localization accuracy over a variable distance.

**Discussion**

In this paper, firstly we carried out a FEM simulation of a Plecotus auritus' ear model at different frequencies and have observed the frequency scanning wherein the lobe direction changes with the frequency. On this basis, we constructed a physical experimental platform in which the active sonar signals were collected. A neural network algorithm was used to estimate the direction of a single target. The experimental results show for single target, this method of bat echolocation detection can achieve good results. A precise estimate of the elevation of a single target is achieved by using pulse train. At last, we design an orthogonal binaural structure for obtaining the aspect angles of two orthogonal directions to perform the target position localization. The results show the varieties of the binaural topological shape can compensate the insufficiency of spatial information.

In our experiments, we mainly make use of the positioning characteristics of Plecotus auritus' auricle at the elevation. The echo features we used are based on frequency sectionalized energy. In the use of neural network for elevation estimation, we also make estimation for azimuth and the results show for single pulse the average estimation probability of ±5° error is less than 50%,as a results, we cannot make use of pulse train to compensate. The reason may be that the energy features of binaural spectrum lack the sufficient phase information needed for the azimuth estimation. The results also prove that for the brown long ear bats, only the elevation direction has a relatively strong correlation with amplitude-frequency energy features of the echoes. For the azimuth estimation, the phase information is important and there have been many literatures focus on the research, for example, mature algorithms such as GCC and MUSIC can achieve the azimuth estimation accurately, but how to successfully combine these methods with characteristics of bat ears to achieve accurate azimuth estimation is also a key direction of our future research.

We can find that the physical experimental results are not constant with numerical simulation results. This might be caused by the following reasons: one is the difference between bat ear physics models and the numerical simulation model. The material of physics models is PLA plastic which is only partial reflection for the ultrasound while numerical simulation model is defined as whole reflection. Second, not like the dot sound source in the numerical simulation, the microphone and the horn have their size and have their own frequency characteristics and space sound field distributions. Their placements and the non-linear characteristics of acoustic wave have also effects for the echo. Third, the effects of the frequency components are not totally equivalent to numerical simulation results.

In addition, there are reverberation and wall surface reflection in indoor space, which can make the receive signals contaminated and hard to be used for the target position. Therefore, how to improve the SNR of the echo is a problem in the echolocation.

There are differences between the experiments and the true bats' behavior. Bats have complicated sonar constructions, and can improve the effects of sonar by many auxiliary ways such as deformations of pinna and movements of lancet. In our experiments, the actual Plecotus auritus' ear is about 30- 34 mm, in order to put an ultrasound microphone into the ear root, the ear model is expanded to the extent of 3 times the size of real models, correspondingly and the frequency used in the experiments is decreased to 1/3 for maintain almost same sound field characteristics. The material of ear model uses PLA plastic, this material physical and biomechanical properties are not comparable to those of real biological materials. This may also affect the reflection and absorption of sound waves. The improved method is to find a smaller ultrasound microphone with higher accuracy, which can be put into a 1:1 ear model; Look for materials similar to real bats' ear tissue in physical and biomechanical properties, which can approach real situation of echolocation of bats.

From the results shown in fig.8,9 and 10, the estimations for the elevation are affected by azimuth angle, but when the scope of the azimuth is limited to - 30° to + 30°, the estimations for the elevation are not only optimal, but also almost no longer affected by the azimuth angle. This can confirm the hypothesis that when the bat catches its prey, firstly it will locate the target's azimuth, its two ears will directly face the target by rotating its head and changing flight direction for getting an accurate estimation for the elevation of the target.

In our experiments, we also found the number of feature parameters extracted from energy of the echo spectrogram can affect the estimation results: first with the increase of the number the estimation effects can be improved in the same data set (training and test data comes from the same data set), but the improvement is limited when the number exceed a certain value; second, the robustness can decreased with the number of feature .This illustrates the excessive features can caused overfitting of the classifier. Apart from the extraction of features, classifier can also bring the difference of the estimation results: we have try the deep learning, cnn,bayes and BP neural network, Their results were similar within same data set but the neural network performed better effects in our robust test.

In this paper, the experiments on the binaural elevation localization of Plecotus auritus were carried out in the laboratory, and the training and testing were carried out on a single target with less interference to the experiment. Therefore, the robustness and generalizability of the system need to be further verified and enhanced. However, this result in the case of single target is a first step towards the application of space target localization by bat ears. The experimental results also prove that the use of bat sonar for localization and navigation still have great development and application prospects, and its more perfect research and application results are worthy of expectation.

**Materials and Methods**

**1 Numeral analyses of the Plecotus auritus ear**

For getting more evidences that inspire us to design good artificial bat-like sonar system, we conducted the numeral simulation for the *Plecotus auritus* ear by using FEM and Kirchhoff integral for the analysis of its spacial Frequency characteristics. The 3D digital model of Brown big-ear was obtained from digital image processing [34] of the Plecotus Auritus ear tomography via CT scanning (Fig. 1). We placed sound source in the inner ear canal of the numerical model and performed FEM numerical calculation.

Finite Element Method and Kirchhoff integral were used for obtaining the beam pattern of the digital ear in the simulation. First the acoustic near field inside a cuboid-shaped volume surrounding the ear were calculated using a finite-element model consisting of linear cubic elements derived directly from the voxel shape representation. Then the far-field directivity pattern was calculated by projecting the complex wave field amplitudes on the surface of the finite-element model's computational domain outwards using a Kirchhoff integral formulation.

**2. Artificial Bat-like device**

The Artificial Bat-like Ear were produced by 3D printer by virtue of the 3D digital representation of the shape of a pinna sample taken from the carcass of a Brown Long-Eared Bat. For avoid damaged in assembling and for fix the microphone on it, the size of the Artificial Bat-like Ear is 3 times as large as the original ear, As the frequency range of Plecotus Auritus ears is 60–20 kHz, according to the scale model principle [35], and the frequency range used in our experiments was adjusted to 15–5 kHz correspondingly. A pair of ultrasonic microphones (SPU0410LR5H-QB) were placed in the inner of a pair of artificial ears and insulating glue was smeared in the gaps between the microphones and the pinna to prevent outside sound waves entering the microphones from bottom of the model, then the artificial ears were fixed to the rotating platform and tilted forward 40°(Fig. 5C). A stepping motor (42BYGH34) was mounted under the rotating platform to facilitate rotation of ear and measure positioning information at azimuth. An ultrasonic loudspeaker (Ultra Sound Gate Player BL Light, Avisoft Company) was fixed under the stepping motor.

### 3. Data Acquisition

All experiments related to the data acquisition were conducted in experimental chamber [8 (L) × 6 (W) × 3.6m (H)] (Fig. 3).No sound insulation were conducted for the chamber. The acquisition parameters are listed in Table 1. The target to be measured was a small ball made of rubber with 11cm diameter suspended by a string. By controlling the height of target, the frequency response of the ear model in different direction could be measured. The ultrasonic signal acquisition and processing device was a signal acquisition card (PXIe-6358 and PXIe-1082, National Instrumental Company, sampling at 100KS/s) which can perform multi-channel synchronous signal acquisition. In our experiments of active echolocation for target, we set the signal acquisition card in the work mode of two-channel synchronous signal acquisition which can gather the binaural signals synchronously.

Ultrasonic loudspeaker emits Linear frequency-modulated pulse signal (Fig. 4A), the detailed depictions of the emitted signals in active echolocation are also listed in Table 1.

### 4.Feature extraction

For obtaining the effective signals for further process, first the endpoint detections were conducted in the signals received by the left or the right microphones, then the effective signals were transferred to time-frequency representation as

$$X_n(e^{j\omega}) = \sum_{m=-\infty}^{\infty} x(m)\omega(n-m)e^{-j\omega m} \qquad (3)$$

Where x(n) is the signal after the process of endpoint detection; and ω(n) is the window function, which shift the sound signal by a step length on the time axis. Here we use hamming windows with a length of 1ms (100 samples) as the window function which shift step is n/2.

The spectrogram representation with 0.5KHz frequency resolution as $|X_n(e^{j\omega})|^2$ of the pulse signals received the left and right microphones are obtained and shown in Fig. 4A.

As spectrogram energy of the chirp pulse signal mainly concentrates near the diagonal line, we set zero of the other parts of the spectrogram to suppress the effects of interference components. Then According to frequencies channels the feature representation of one spectrogram is restricted to a 30, and the left and the right echoes signals form one vectors with 60 elements for the input of classifier.

**5 Neural network Classification in active echolocation**

The received pulse Spectrograms are complex patterns which are assumed to carry important information to discriminate among different locations of the target. The classification and direction estimation task is considered as a pattern recognition problem. Here a standard BP neural network which consists of an input layer with 60 neurons (30+30 i.e.,the extracted features from the signals of left and right artificial ears that are directly fed into the network) a hidden layer with 9 neurons, and an output layer with 1 output neuron is applied for the directions of target estimation. The three layers in the BP neural network are fully connected. The output neuron represents azimuth or elevation angle of the target respectively.

**6. Moving windows accumulation discrimination by virtue of pulse series**

The activities of the output neurons indicate the respective value of the azimuth or elevation in an analysis of one pulse. If multiple pulses for one target were analyzed by the same classifier, multiple values of the estimation will be outputted. Usually from the multiple values of the estimation an optimization method can be find and optimization value as the result of the method can be come out which located at a higher level of confidence. This principle is suggested why the bats locates the target by virtue of pulse train [36][37]. For imitating the actual signals which are often emitted by the bats in the form of a pulse train, we designed the improved direction angle discrimination method named moving windows cumulative discrimination based on pulse train. The moving windows cumulative method can get an optimal estimation using multiple relatively rough estimate values. For n initial estimate values, A P-levels moving windows cumulative method is depicted as follow:

For n estimated values (each value represents one sample) obtained from a certain method, such as BP Neural network, the first level moving window with length L moves over the range of all the n samples with a step length l. The starting position of the window (left edge) is aligned with the point with minimum value, and the ending position (right edge) is the sample point with maximum value. Each point in the range was set an initial token value y as 0. Whenever the window moves one step, the y value of each point in the window increases by the number of the samples in the current window. When the window moves to the ends, we calculate the point with the maximum y value as the optimal value of first-order moving window. If there are more than 1 point with equal maximum value, take intermediate point of the left and the right point with maximum value as the optimum solution. Then an L/2 length second-level window and L/4 length third-level window estimation can also be performed up to a pth window estimation. The number of levels determines the accuracy of the estimation. Usually, the accuracy of the estimation is less than the half of the window's width.

## 7. Cross-validation

In our active echolocation experiments (section 2.1), the Artificial Bat-like Ear device Emits and receives 200 pulses at 8 sites surrounding the spherical target(fig 10).So we can select 10 group pulse as 10 pulse trains each of which contains the same number of pulses. We conducted five cross-validations in each of which the number of pulses in one pulse train is 3,5,10,15,20  (fig.7). For obtaining more reliable results and to suppress effect of overfitting, tenfold cross-validation is used to test the reliability of the neural network system along with the established angle recognition system. The detailed process is as follows: 10 combinations were analyzed, each one including a training set of 9×8 recorded pulse trains (9 pulse trains for each site )  and a testing set of 1×8 recorded pulse trains (1 pulse trains for each site) for one target locations to be estimated.

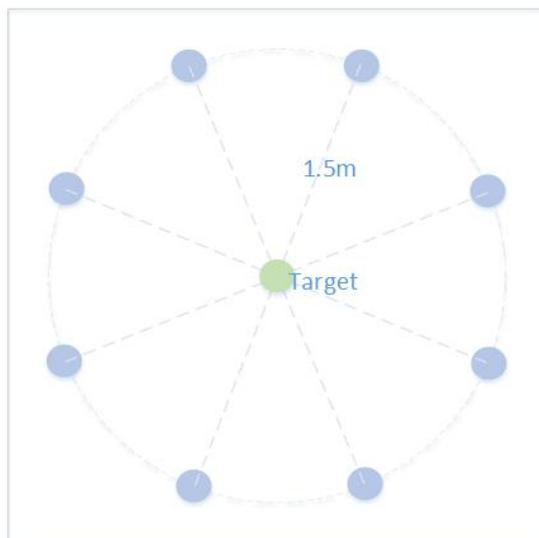

**Fig.10.The 8 different sites surrounding the spherical target used for locating the Artificial Bat-like Ear device for the experiments in section 2.1 and the training data acquisition in section 2.2.**

highly diverse foraging and echolocation behaviors of microchiropteran bats.Front. Physiol., 03 July 2013　https://doi.org/10.3389/fphys.2013.00164

**Acknowledgments**

Funding: This work was supported by the National Natural Science Foundation of China for from Grant No. 61271453, the Shandong Provincial Key R&D Program 2017GGX10113, Shenzhen science and technology research and development funds (No.JCYJ20170818104011781), the National Natural Science Youth Foundation of China for from Grant No. 11704154.

Author contributions

Sen Zhang: Detailed simulation and hardware experiment, and paper writing.

Xin Ma: The general idea of the article, experimental guidance and financial support.

Hongwang Lu: Provide ear model of Plecotus auritus and numerical simulation calculation.

Weikai He: Print 3D model, prepare experiment equipment and guide part of experiment.

Weidong Zhou: Guide neural network theory.

Competing interests

The authors declare no competing interests.

Correspondence and requests for materials should be addressed to Xin Ma.

(max@sdu.edu.com)